# Identifying OSPF Anomalies Using Recurrence Quantification Analysis

Bahaa Al-Musawi* and Philip Branch†
*Faculty of Engineering, University of Kufa, Al-Najaf, Iraq
†School of Software and Electrical Engineering, Swinburne University of Technology, Melbourne, Australia
{balmusawi,pbranch}@swin.edu.au

***Abstract*—Open Shortest Path First (OSPF) is one of the most widely used routing protocol to manage intra-domain routing. OSPF has been identified with many serious security issues. LSA falsification is one of the most critical vulnerability that can cause route loop and black hole. Network operators need to rapidly identity such anomalies. Network operators need also to identify hardware failure. In this paper, we investigate the capability of Recurrence Quantification Analysis (RQA), an advanced non-linear statistical analysis technique, to identify OSPF anomalies. We evaluate the capability of RQA to identify OSPF anomalies using a controlled testbed where we introduced different types of LSA falsifications as well as hardware failures. Our evaluation shows that RQA can rapidly detect OSPF anomalies.**

***Index Terms*—Intra-domain routing, OSPF, anomaly detection, RQA, testbed, CORE.**

## I. Introduction

Open Shortest Path First (OSPF) has been designed to be deployed within a single an Autonomous System (AS) where an AS represents a large organisation or an Internet Service Provider (ISP). OSPF is one of the most widely used interior gateway protocols [1], [2]. It is a link-state routing protocol where each router maintains a database that describes the AS's topology [3]. The main responsibility of OSPF routing protocol is to allow all routers within an AS to construct their routing tables and updates them when a change in the AS's topology occurs.

OSPF supports authentication where each link between OSPF routers is associated with a shared key. However, OSPF has been identified with a serious security issues [4]. Link State Advertisement (LSA) falsification is the most well-known security issue. LSA falsification occurs when an attacker sends a false LSA on behalf of the victim router which can cause a serious impacts such as routing loop and black hole. Although OSPF provides "fight-back" mechanism by instantly sending LSA when OSPF routers receive a false information in the LSAs which originated by them, different types of LSA falsification overcome this mechanism [5], [1], [6]. Monitoring such serious attacks and hardware failures is a highly demanding for effective management and operation of IP routing protocol [2]. This paper introduces a new approach to detecting OSPF attacks and hardware failures which we define them as OSPF anomalies.

Our approach is based on using Recurrence Quantification Analysis (RQA) method. RQA is a non-linear statistical analysis method uses the concepts of phase plane trajectory [7]. RQA is a tool to extract hidden information from statistics of dynamic non-linear systems [7]. To evaluate the ability of RQA to detect OSPF anomalies, we use OSPF traffic obtained from a controlled testbed where we introduced different types of OSPF anomalies. Our evaluation shows the ability of RQA to instantly identify different types of OSPF anomalies.

The rest of this paper is organised as follows. Section II introduces a brief background of OSPF and different types of LSA updates. Section III outlines our approach using RQA. In Section IV, we introduce our testbed and discuss the selection of monitoring point to capture OSPF traffic. We evaluate our approach to detecting OSPF attacks in Section V and finally conclude our work in Section VI.

## II. OSPF Background

The Internet is a decentralized global network comprised of tens of thousands of Autonomous Systems (ASes). An AS is a set of routers under a single technical administration using an Interior Gateway Protocol (IGP) such as Open Shortest Path First (OSPF) to communicate with other routers within the AS and an Exterior Gateway Protocol (EGP) such as Border Gateway Protocol (BGP) to communicate with other ASes [8]. OSPF is one of the most widely used IGP. OSPF was firstly described in RFC1131 and revised by RFC 2328 [3]. There are two versions of OSPF, OSPF v2 supports IPv4 described in [3] and OSPF v3 that supports IPv6 described in [9].

OSPF is a link state routing protocol where each OSPF routers advertise the state of their links to their neighbor routers. When an OSPF protocol is enabled for a particular link, information associated with that router is added to the local Link State Database (LSDB). Afterward the router sends Hello messages on its operational links to determine whether other link state routers are operating on the interfaces as well. In addition to neighbor discovery purpose, Hello messages are sent to maintain adjacencies between neighbor routers. After establishing the adjacency between two OSPF routers, these routers exchange a summarised LSDB. Each OSPF router compares the received summary with its local LSDB to ensure it is up to date. If one routers realize that it requires an update, it will request the new information from the adjacency router.

OSPF instantly sends Link State Advertisement (LSA) messages to reflect a change in the topology and every 30 minutes to refresh their routers database. These LSAs messages are disseminated throughout the entire OSPF domain. There are five LSA types which are summarised in Table I.

TABLE I
A SUMMARY OF LSA TYPES

| LSA type | Description |
|---|---|
| Type-1 | Router LSAs which describe the states of router's interfaces |
| Type-2 | Network LSAs which describe the set of routers attached to the network |
| Type-3 | Summary LSAs which describe routers to networks |
| Type-4 | Summary LSAs which describe routers to AS boundary routers |
| Type-5 | AS-external LSAs which describes routes to destinations external to the AS |

OSPF is a hierarchical routing protocol by supporting sub-domains or areas. Dividing one domain into areas limits the scope of LSAs flooding within the OSPF domain. OSPF supports authentication and encryption. It also provides "fight-back" mechanism when a router receives a false LSA that was advertised by another router on its behalf, the router immediately advertises a newer instance of the LSA which cancels out the false one. However, OSPF has several known security issues [1], [4], [6].

LSA falsification is the most critical security issue in OSPF. It occurs when an attacker advertises an LSA with false link information. False LSAs can be injected into the network from a subverted router or common network host. If a false LSA is accepted by at least one router, it allows the attacker to poison the routers' view of the AS topology and hence affecting its routing table [4]. LSA falsifications are classified into self-LSA falsification where a malicious router falsifies only its own LSA and other–LSA falsification where the malicious routers advertise a false LSA on behalf of other routers [**?**]. Early detection of OSPF attacks enables network operators to protect their network from worst consequences such as denial of service, eavesdropping and network delays. In addition to OSPF attacks, rapidly detection of router and link failure in an OSPF domain is of interest to network administrator. In this paper, we refer to LSA falsification and hardware failures as OSPF anomalies. In next section, we introduce our approach to detect OSPF anomalies using a non-linear statistical analysis technique.

## III. A NON-LINEAR APPROACH TO IDENTIFY OSPF ANOMALIES

Different methods and models for analysis of time series and forecasting are available. The Fourier transform (FT) and the Auto-Regressive Integrated Moving Average (ARIMA) model are the most well-known for analysis and forecasting time series. The FT and the ARIMA have some limitation for analysis of time series. The FT has some limitations such as time and frequency positions, non-stationarity or abrupt changes in a signal can spread out for whole signal as well as resolution. These drawbacks limit the application of the FT in detecting short periods of anomalous OSPF behaviour. The ARIMA model has two significant limitations: (1) future values are assumed to be a linear function of past values and (2) a large amount of historical data is required to obtain reliable predictions [10].

Here, we use Recurrence Quantification Analysis (RQA), a non-linear technique based on a phase plane trajectory, to identify OSPF anomalies. RQA was introduced to quantify the important aspects revealed by Recurrence Plot (RP), a graphical method to display recurring patterns and non-stationarity in time series.

### A. Recurrence Plot

Recurrence Plot (RP) is a tool to visualise the time-dependent behaviour of the dynamics of a system using the concepts of phase plane trajectory, and simplifies interpretation of recurrent data. With enough data, structural patterns in the RP can reveal information about the time evolution of the phase space. RP is not limited to long data sets. It can be used for short, noisy, and non-stationary data sets [11]. RPs can be formally expressed by the matrix $R$

$$\mathrm{R_{i,j}}\left(\varepsilon\right) \quad = \quad \Theta\left(\varepsilon - \|\overrightarrow{x_i} - \overrightarrow{x_j}\|\right),\ i,j = 1\,,\ldots,\,N, \quad (1)$$

where $R_{i,j}$ is an element of the recurrence matrix $R$, N is the number of measure points, ε is a threshold distance, $\Theta\left(\cdot\right)$ the Heaviside function and $(\|\cdot\|)$ is a normalization operation.

To construct an RP, three parameters have to be carefully selected. These are time delay $(\tau)$, embedding dimension $(m)$, and the threshold $(\varepsilon)$. Selecting non-optimal values for RP's parameters can produce different structures for the same input data. For example, non-optimal values of embedding parameters can cause many interruptions to the Line of Identity (LOI), a black main diagonal line with an angle $\frac{\pi}{4}$ in the RP.

To estimate time delay, the Auto-correlation function (ACF) and Mutual Information (MI) are the most well-known methods to determine time delay. Unlike ACF which measures linear correlation, MI measures both linear and non-linear correlation. Therefore, we will use the MI method to determine the time delay parameter. The first minimum value of MI represents the value of time delay. For a periodic data, the value of periodicity can be used as the value of time delay. The embedding dimension parameter can be estimated using False Nearest Neighbour (FNN), a tool for determining the proper embedding dimension in dynamic systems. The FNN requires the value of time delay which should be calculated first. Once again, the first minimum value of FNN represents the value embedding dimension. Although there is not a well-established method to determine the optimal values of the threshold, the value of threshold has to be selected to be as small as possible. A recommendation from [7] suggests that the threshold has to be selected less than 10% of the maximum phase space diameter.

### B. Recurrence Quantification Analysis

Interpreting an RP requires a high level of experience especially for complex data. To reduce the difficulty of RP interpretation, RQA has been introduced. In addition, RP cannot be directly used for automated detection of system behaviour changes or real-time anomaly detection. RQA has been previously used to detect instability and anomalies in different disciplines such as physics, medicine, and engineering [10], [12].

RQA provides several measures of complexity which are described in Table II. These measurements demonstrate the characteristics of systems at different times. For example, RR refers to the probability that a system recurs after a number of time states while TT can be used to measure how long the system remains in a specific state [11]. Our approach is based on on the concept that a significant change in RQA measurements represents an indication for anomalous behaviour in the underlying system behaviour that identify anomalies.

TABLE II
A DESCRIPTION OF RQA MEASUREMENTS

| RQA measurements | Description |
|---|---|
| RR | Measure the density of recurrence points in the RP |
| DET | A measure based on diagonal lines of the RP |
| TT | Contains information about the vertical structures in the RP |
| T2 | Measure the average lengths of the white vertical lines in the RP |
| ENTR | It is the Shannon entropy of the frequency distribution of the diagonal line lengths |
| L-MAX | A measurement that is based on diagonal lines of the RP |
| L-MEAN | Measure the average diagonal line length in the RP |

## IV. Experimental Setup

The Common Open Research Emulator (CORE) is a real-time network emulator. It supports routers, hosts and simulates the network links between them [13]. CORE can provide a realistic running of emulated networks with a relative inexpensive hardware. CORE combines the ability of simulation tools such as ns-3 [14] and emulation tools such as PlanetLab [15] by emulating the network stack of routers or hosts through virtualization, and simulating the links that connect them together.

In this paper, we use the CORE in our experiment and evaluation where the routers were configured to use OSPF v2 with equal weights. We also use tcpdump, a Linux tool to capture packets on a network interface, and Scapy, a Python-based powerful interactive manipulation framework which is able to decode a wide range of protocols [16], to filter OSPF traffic.

### A. Selecting Monitoring Point

OSPF routers send a periodic refresh of LSAs, even when there is no change in the topology. The default value of the refresh period is 30 minutes. In addition, OSPF routers flood LSAs when there is a change in the network topology. For example, when a link between two routers comes down, the two routers have to originate and flood their LSAs with a new link included in it. Although LSAs updates disseminate through the entire OSPF domain, selecting a monitoring point to capture LSA updates needs care.

In OSPF domain, networks can be classified into transit and stub network. Transit network send and receive LSA updates with other networks while stub network receive only LSA update. Using OSPF traffic captured from transit network might not reflect the actual behaviour. This can occur when two OSPF routers within same area send an LSA update to each other which are identical except LS age, an OSPF field that specifies the age of the LSA in second. It is set to 0 when the LSA is originated and incremented on every hop of the flooding as well as they are held in each router's database [3]. In such a scenario, the LSA which have the smaller age value should be accepted by the received router and this router should acknowledge this LSA by sending an acknowledgment update. To clarify such a scenario, consider Figure 1. In this figure, R1 and R2 are two adjacencies OSPF routers. In this scenario, there are three cases that can described the exchanging of OSPF updates between R1 and R2.

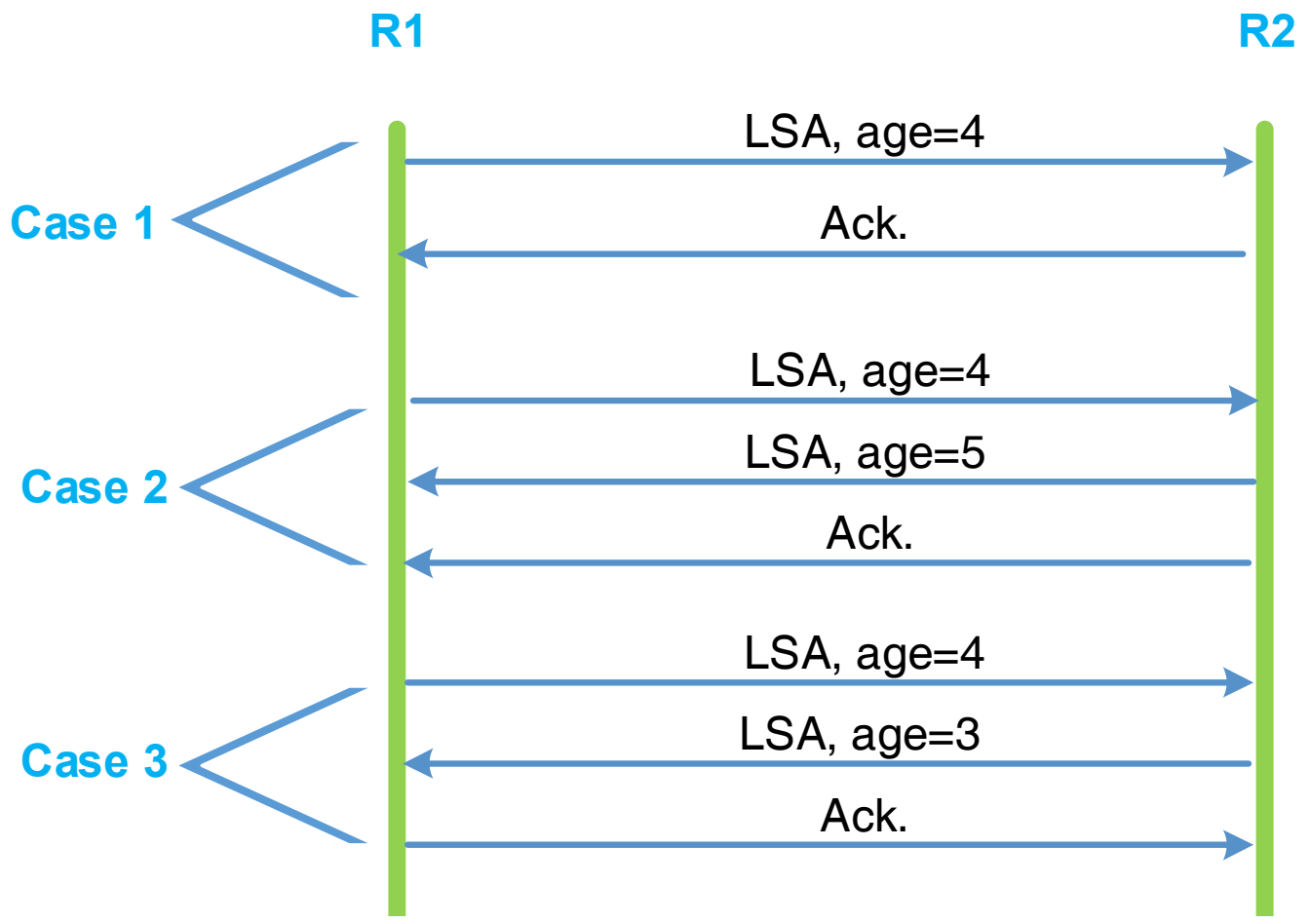

Fig. 1. An example of sending OSPF traffic between two nodes

*Case* 1. R1 or R2 sends an LSA update and the other router acknowledges it. Such a case can be seen in stub routers.

*Case* 2. R1 sends an LSA with age 4 and R2 sends the same LSA but with age 5. Although R2 sent LSA after R1, R2 has to acknowledge the LSA sent by R1 as it has a smaller age value.

*Case* 3. R1 sends an LSA with age 4 and R2 sends the same LSA but with age 3. R1 discards the LSA that sent and will acknowledge the LSA sent by R2 as it has a smaller age value.

As we can see from case 2 and 3, there are extra information sent between the two nodes. This can produce inaccurate representation for system behaviour. Therefore, we will monitor OSPF traffic by using LSA updates sent by stub network.

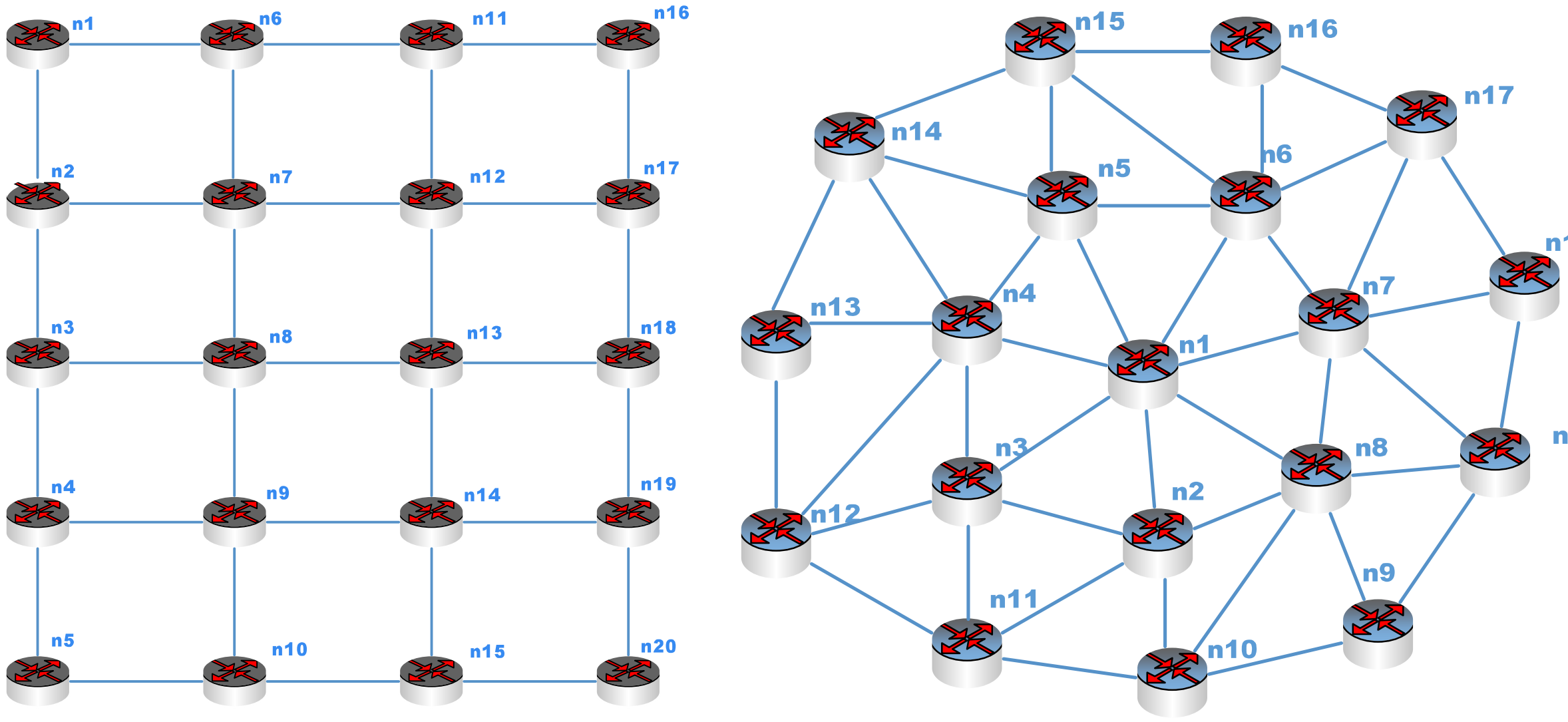

(a) Topology-1

(b) Topology-2

Fig. 2. Investigating the effect of topology and number of peers on RQA parameters

### *B. Estimating RQA parameters*

In this section, we explore if the estimation values of RQA parameters (time delay, embedding dimension, and the threshold) depends on the topology and number of OSPF routers for an OSPF network. In other words, does the topology and number of OSPF routers in a single OSPF area can change the characteristics of RP for OSPF traffic? For that purpose, we adopt different topologies. In each topology, we monitor OSPF traffic on different routers. The purpose of this investigation is to find out if we can use (generalise) RQA parameters for any topologies or we need to find out these parameters for every topology.

A recommendation from Cisco that number of OSPF routers in a single OSPF area should have no more than 50 routers and for areas that suffer from unstable links should be even smaller [17]. Therefore, we will use 20 routers and 35 routers in our adopted topology taking in our consideration different number of peers per OSPF router. Figure 2 show examples of two used topologies. For example, in topology 1 the maximum number of peers is 4 while in topology 2 is 6. Whatever the number of OSPF routers and number of peers per OSPF router, we can see that the characteristic of OSPF traffic per OSPF router shows similar RQA measurements ($\tau = 1$, $m = 2$, and $\varepsilon = 0.2$).

## V. Evaluation

To evaluate the capability of RQA to identify OSPF anomalies, we apply our RQA approach to OSPF traffic collected from a control testbed. RQA measurements change to reflect changes in the underlying system traffic [18]. Our approach is based on an assumption that a significant change in one of RQA measurements represents an indication for identifying OSPF anomalies.

We use two types of anomalies. These are a hardware failure, link or a node failure, and different types of OSPF attacks. In both type of anomalies we use the topology shown in Figure 3. We start our evaluation using a hardware failure.

### *A. Hardware Failure*

In this section, we introduce multiple link failures by shutting down multiple interfaces on multiple OSPF routers and capture LSA updates from multiple monitoring points. To get an accurate evaluation in term of time and sequence, we build an automation script that introduces multiple failures on routers. These routers are abr1 and r6 which have been selected based on their locations in the OSPF area. We start our experiments by shutting down interface eth0 of abr1 and ending with turning up eth0 of abr1 and eth1 of r6 as shown in Figure 4. The time delay between each event is 4 hours (14400 seconds). We capture OSPF traffic from multiple monitoring points including rcs1, r7, r11, r12, r13, r14, r15, and r16 and originated by abr1. Figure 5 shows number of all LSAs and LSAs type-1 per second captured from multiple monitoring points.

Although OSPF updates have captured from stub networks, the total number of all LSA updates and LSA updates type-1 per second are not identical. This is due to LSA updates arrive to the monitoring points at different time stamps. For example, Figure 6 shows number of all LSA updates captured by OSPF routers rcs1 and r16 where we can see how LSA updates arrived at different time stamps. However, the total number of LSA updates captured at all multiple monitoring points are equal.

Furthermore, the number of all LSAs and LSAs type-1 captured from OSPF router r14 shows different behaviour where there is no OSPF updates during the period 72001-86402 seconds. This is due to the introduced hardware failure at time stamp 72000 seconds. As a result of losing connection between r14 and r6, there were no LSAs during the period 72001-86402 seconds. We can also see that at time 86403 seconds that there are 3 LSAs captured by r14 and advertised by abr1. The extra one LSA is due to when the adjacency between r6 and r14 going up, r14 sends a link state request packet. This packet is sent to request the pieces of the r6's database that are more up to date [3].

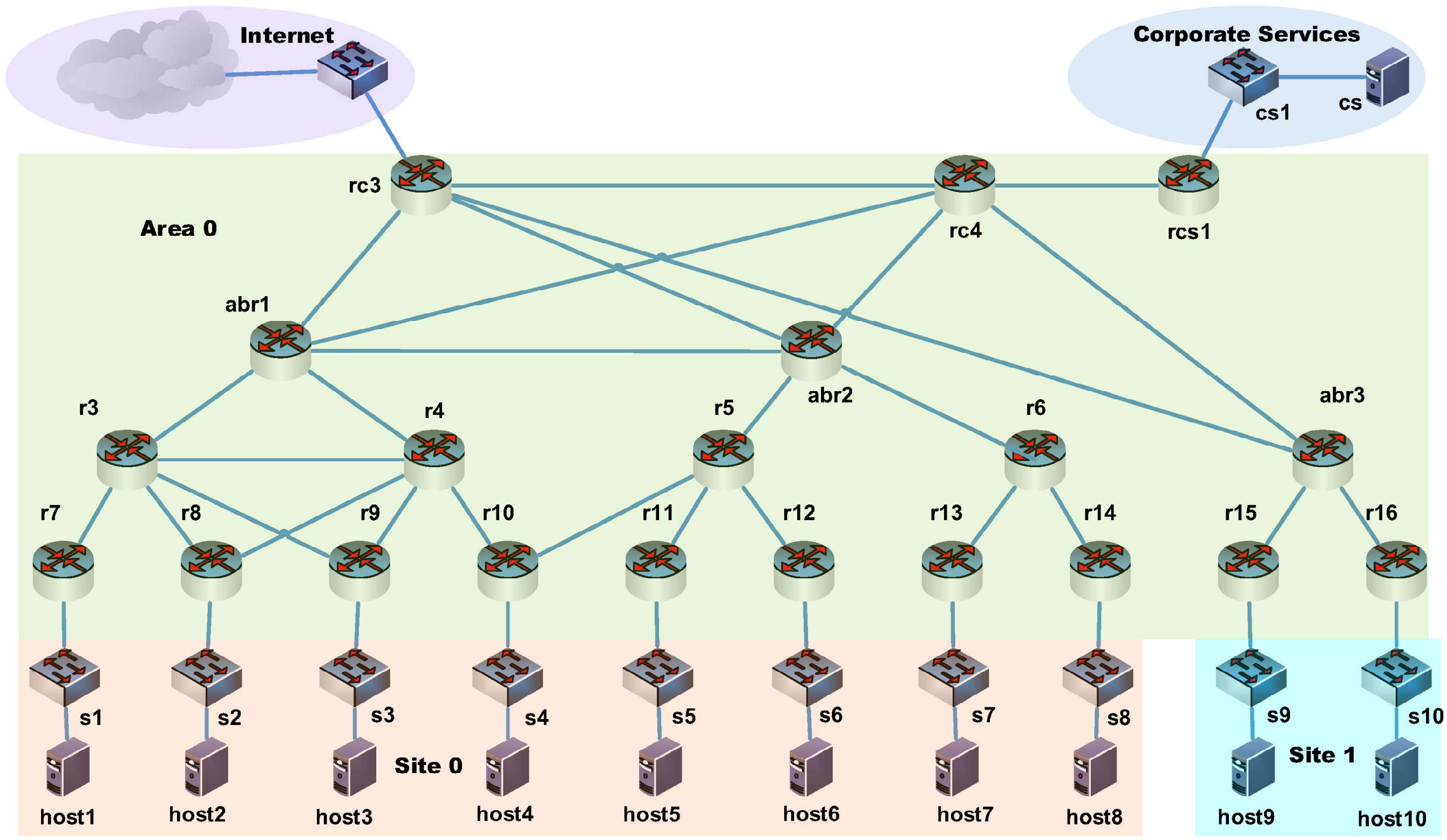

Fig. 3. An example of sending OSPF traffic between two nodes

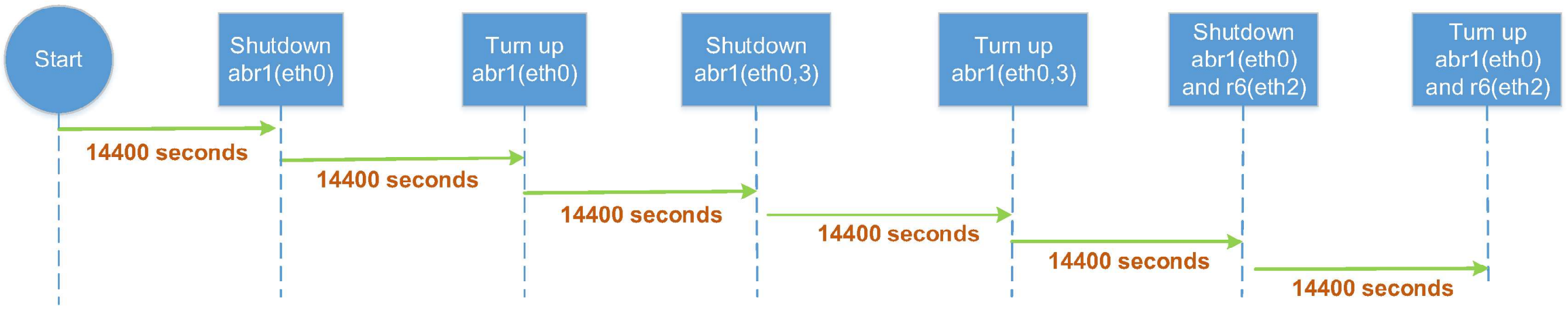

Fig. 4. An example of sending OSPF traffic between two nodes

Before calculating RQA measurements for the captured OSPF traffic, we sample LSA updates every 10 seconds as a bin size. RQA measurements then calculated every bin using 200 bins as a history. Figures 7 and 8 shows number of all LSAs and LSAs type-1 captured from rcs1 and advertised by abr1 and their corresponding values of RQA measurements respectively. RQA is able to identify hardware failure within 2 bins size (20 seconds) as shown in Figure 9. Although RQA shows its ability to rapidly identify interfaces shutting down on different routers as well as turning up multiple interfaces on different routers, it does not able to identify multiple shutting down for multiple interfaces on different routers, the event at time stamp 7200 bins.

### B. OSPF Attacks

In this section, we evaluate the ability of RQA to detect the most well known OSPF attacks. Partitioning attack [6], disguised attack [5], and adjacency spoofing attack [1] are the most well known OSPF attacks. Partitioning attack is an example of self-LSA falsification attacks while disguised and adjacency spoofing attacks are examples of other-LSA falsification. The consequences of these attacks can cause routing loop, long routes, routers become disconnected, and link overload as a result of diverting large volume of traffic through a limited capacity link [5], [6].

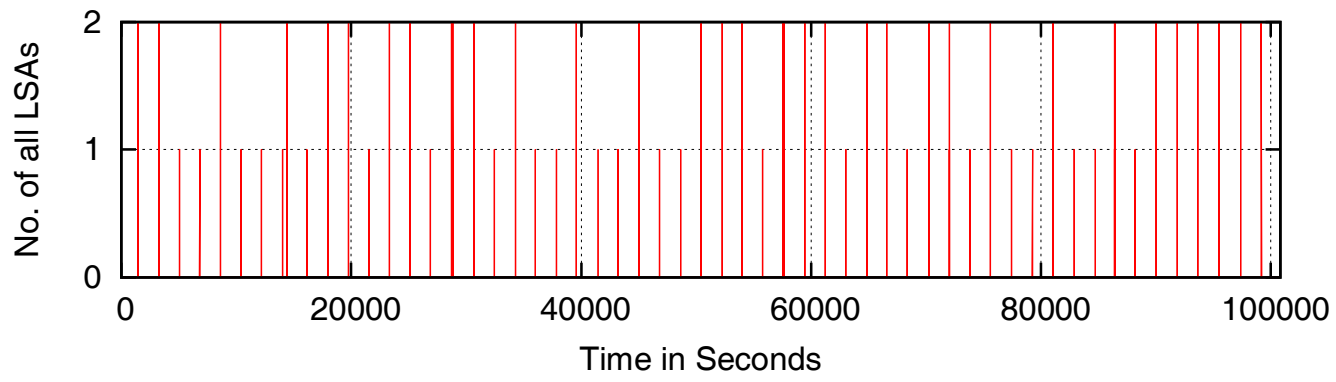

(a) Number of all LSAs updates captured by rcs1

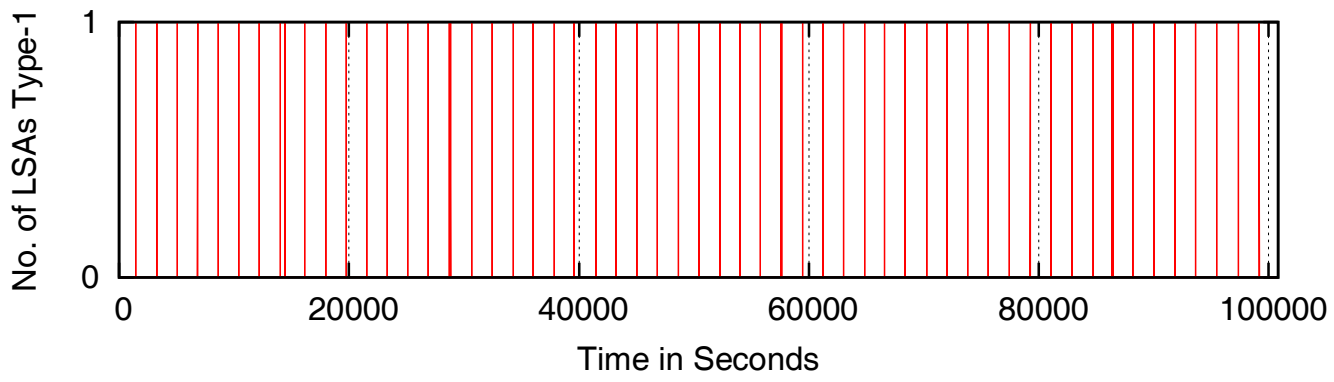

(b) Number of LSAs updates type-1 captured by rcs1

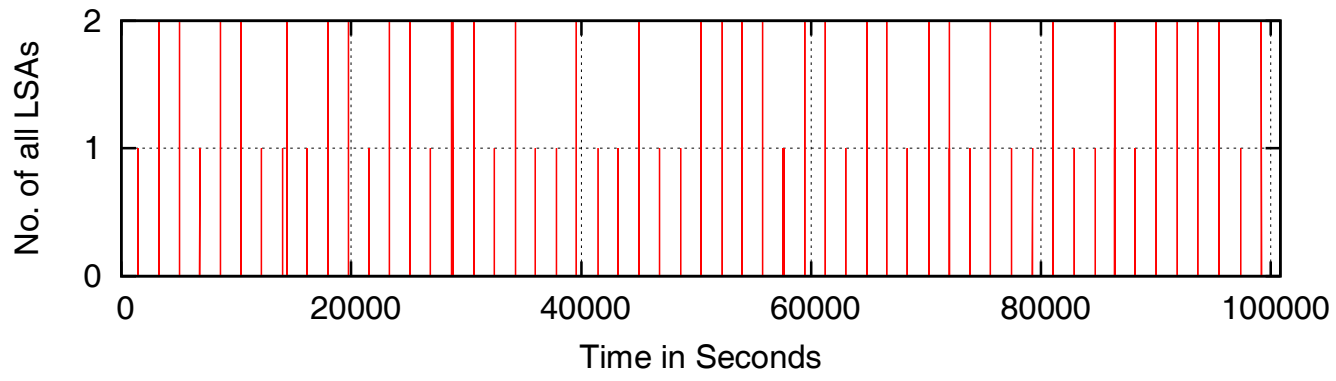

(c) Number of all LSAs updates captured by r16

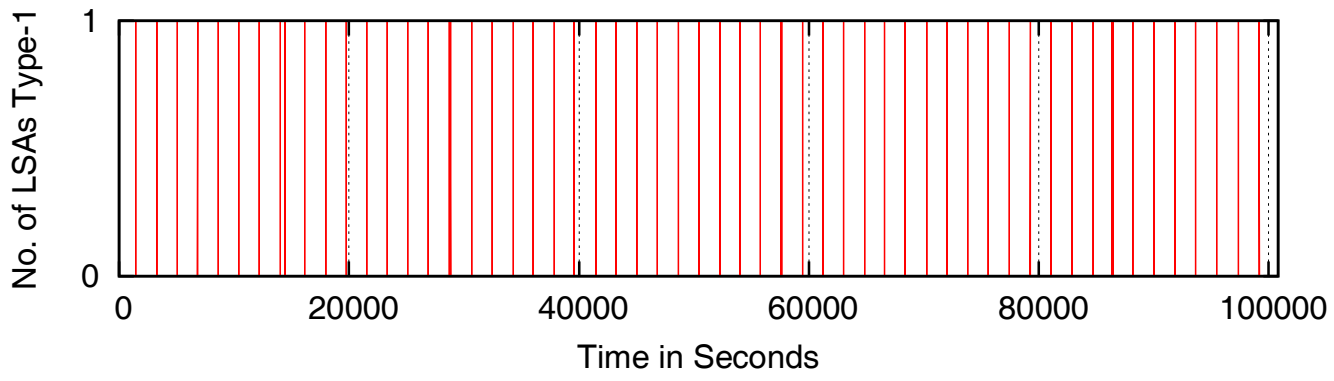

(d) Number of LSAs updates type-1 captured by r16

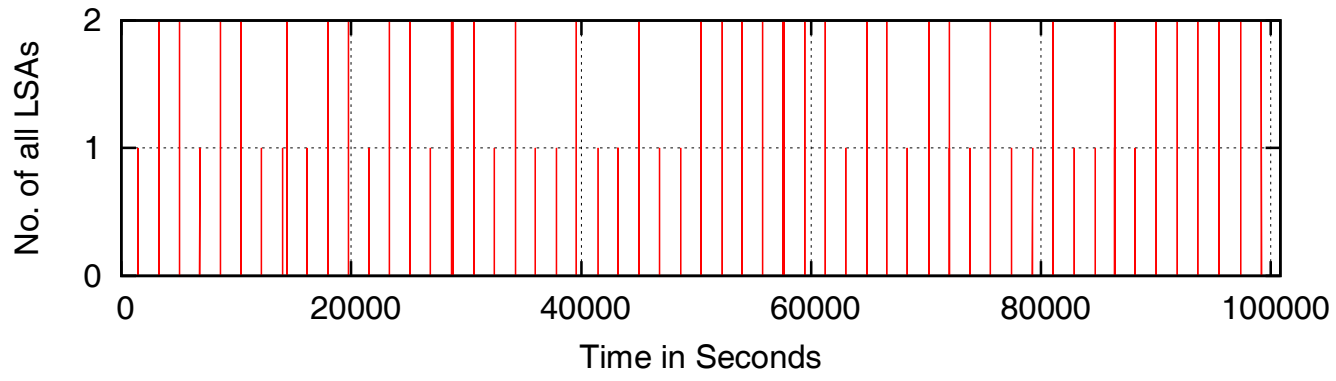

(e) Number of all LSAs updates captured by r15

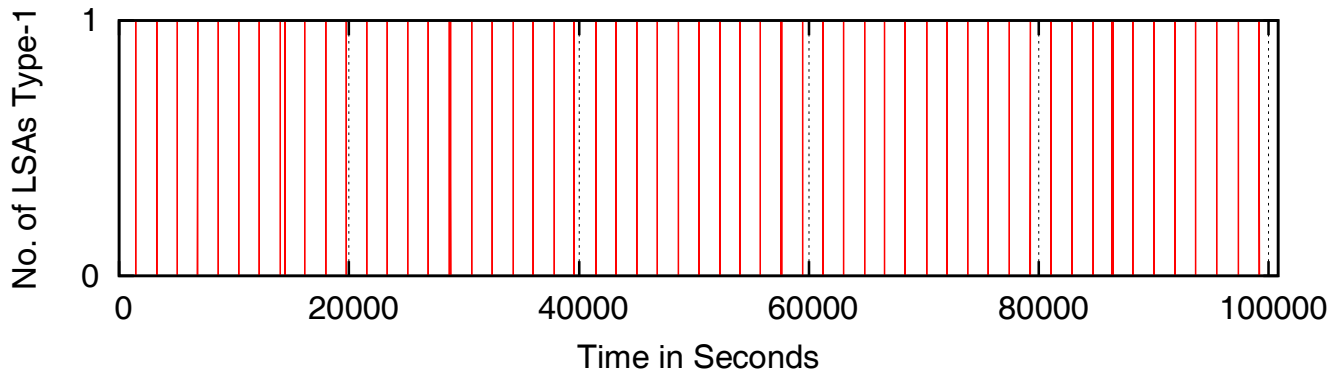

(f) Number of LSAs updates type-1 captured by r15

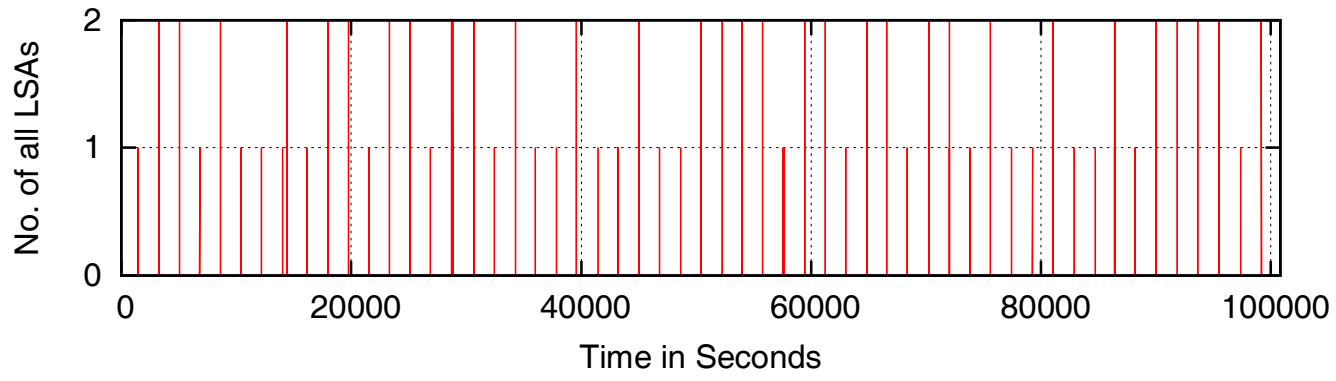

(g) Number of all LSAs updates captured by r7

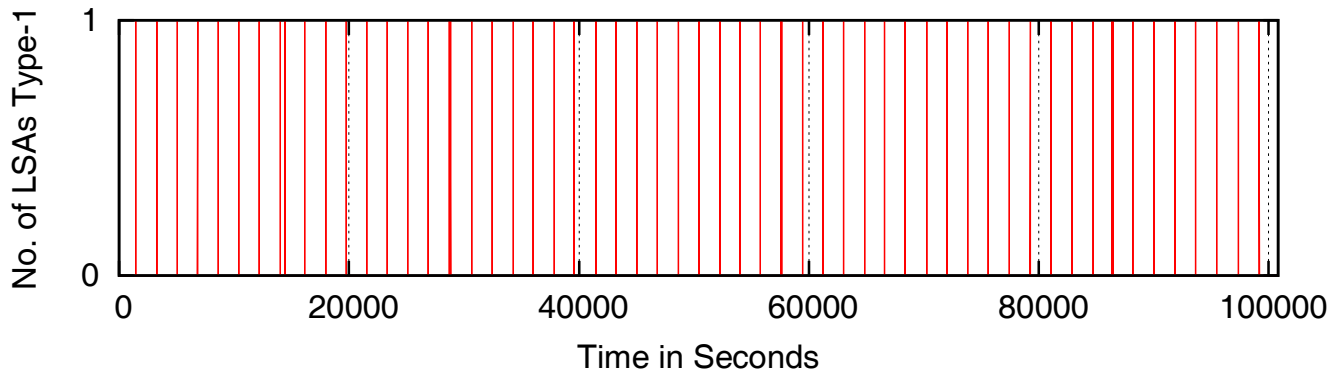

(h) Number of LSAs updates type-1 captured by r7

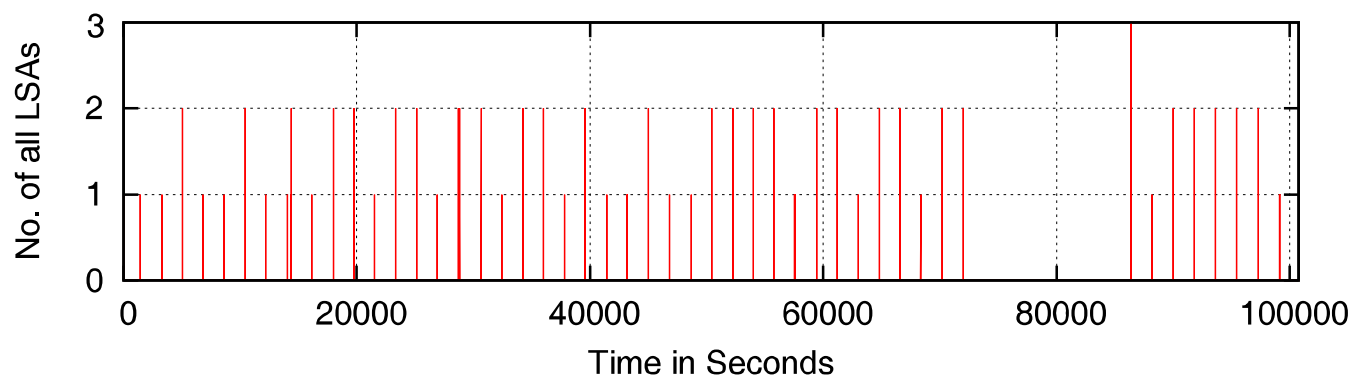

(i) Number of all LSAs updates captured by r14

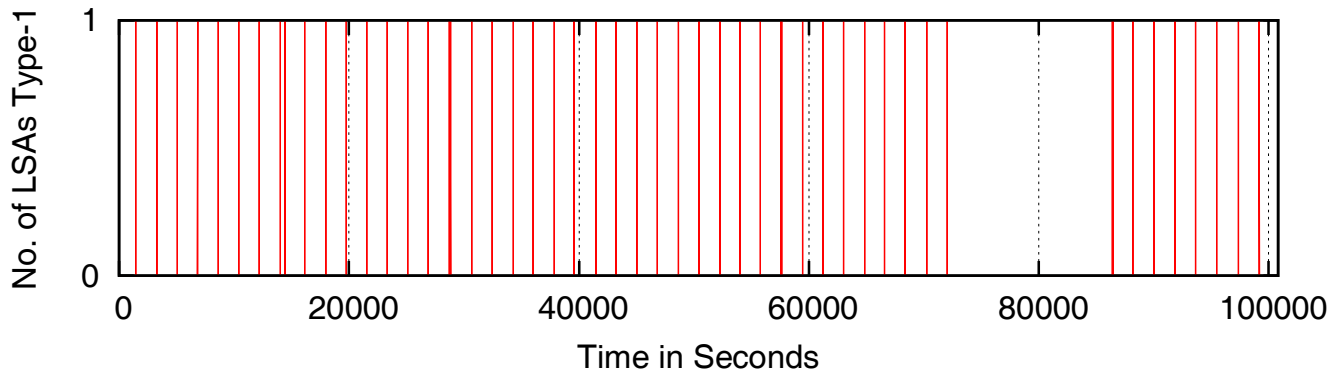

(j) Number of LSAs updates type-1 captured by r14

Fig. 5. Number of all LSAs and LSAs type-1 captured by different monitoring points

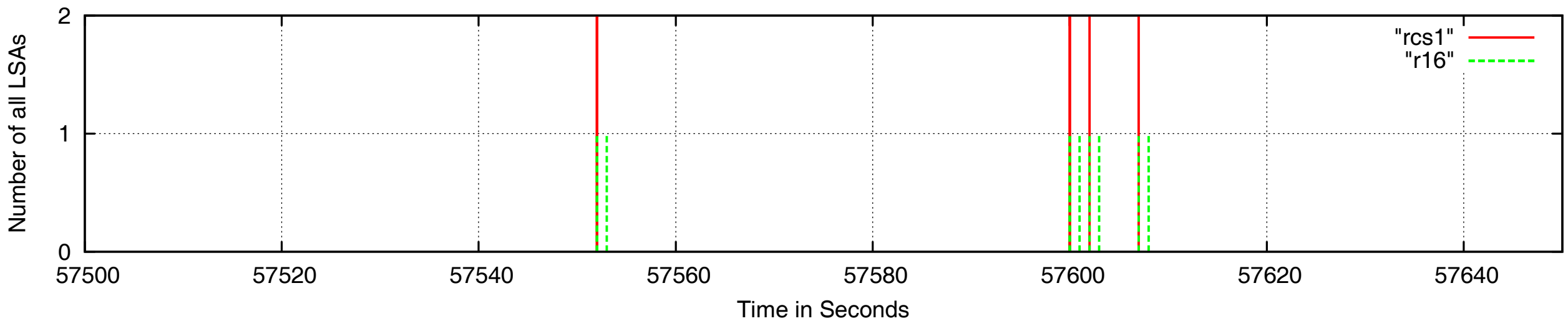

Fig. 6. An example of arriving LAS updates to monitoring points at different time stamps

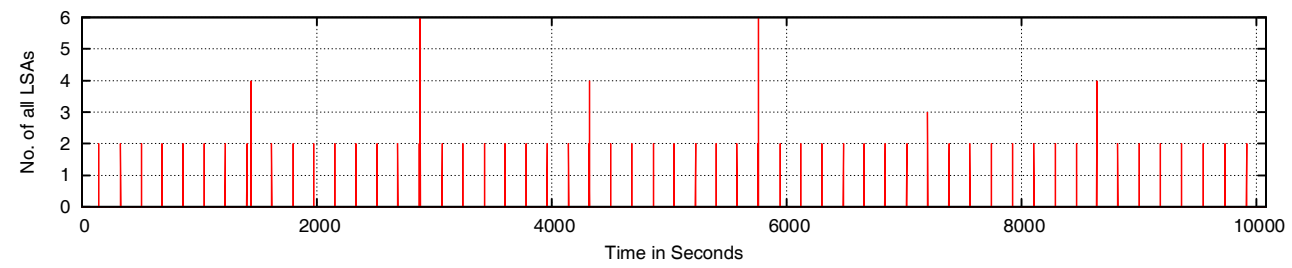

(a) Number of all LSAs captured by rcs1

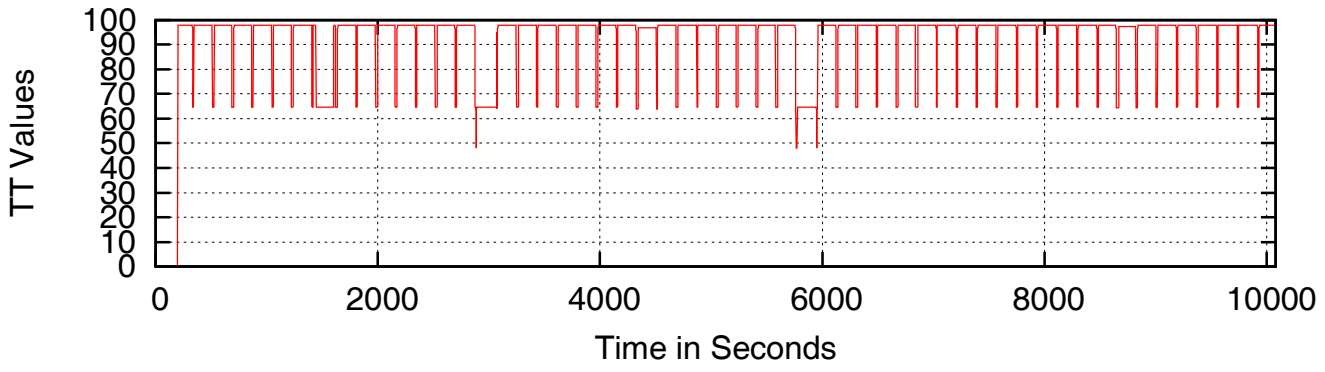

(b) TT measurement

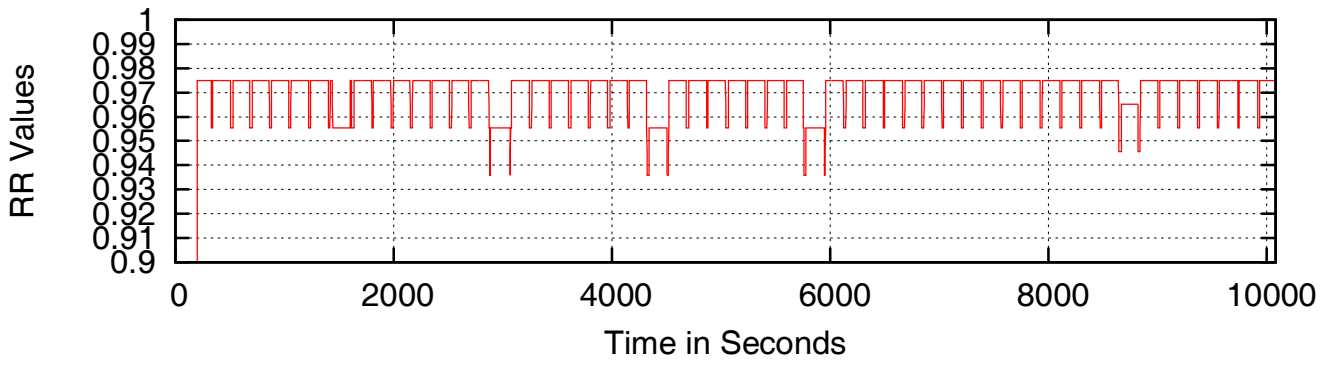

(c) RR measurement

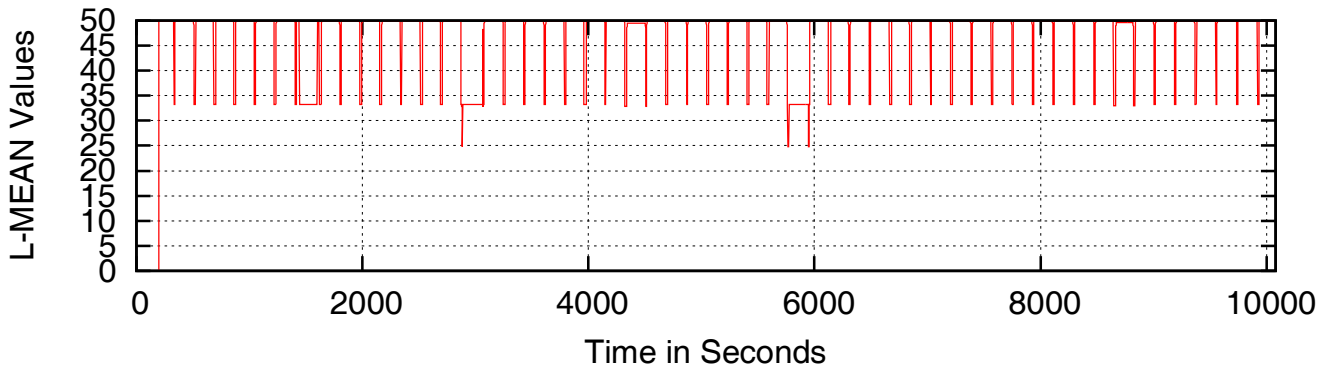

(d) L-MEAN measurement

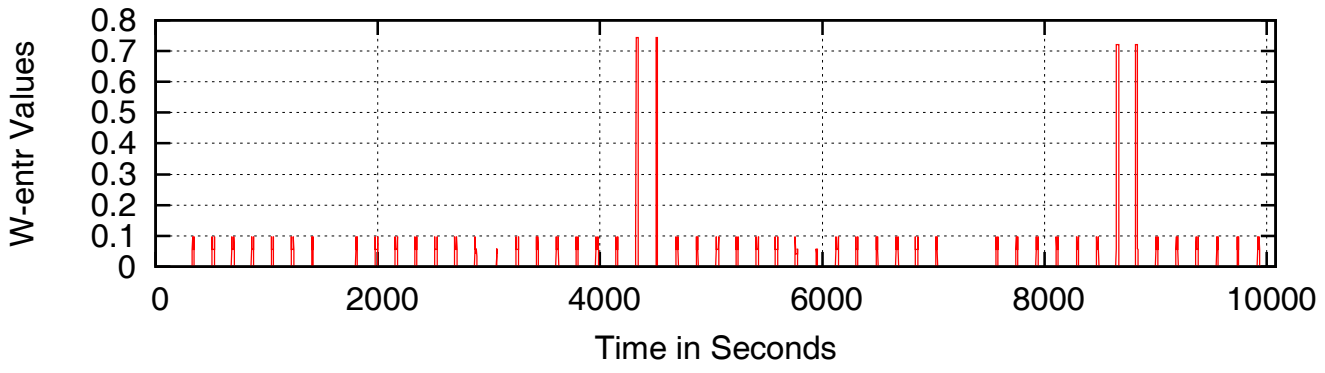

(e) W-entr measurement

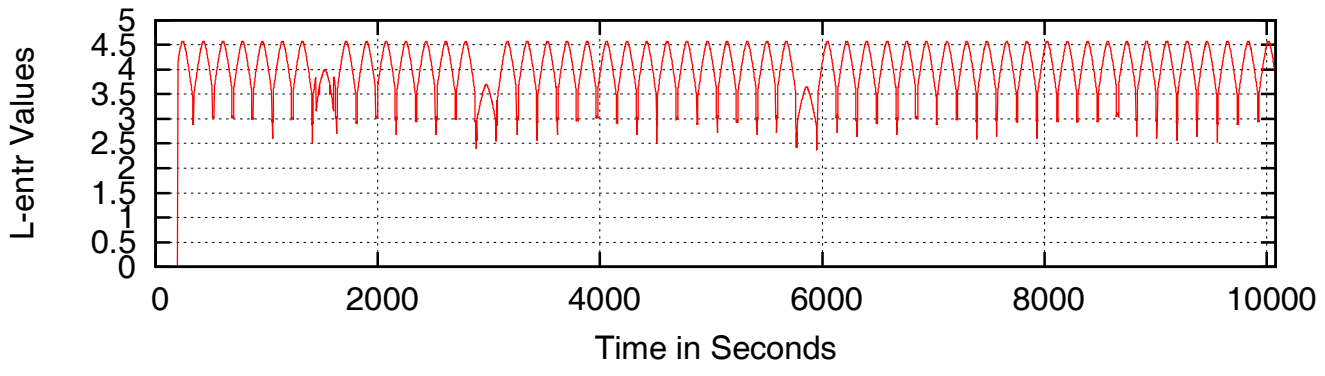

(f) L-entr measurement

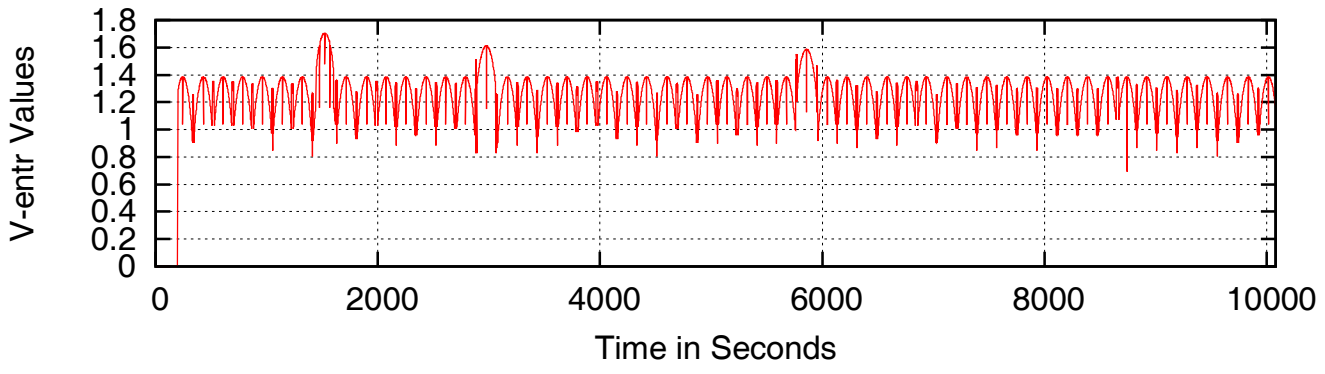

(g) V-entr measurement

Fig. 7. Number of all LSAs advertised by abr1 and its corresponding values of RQA measurements

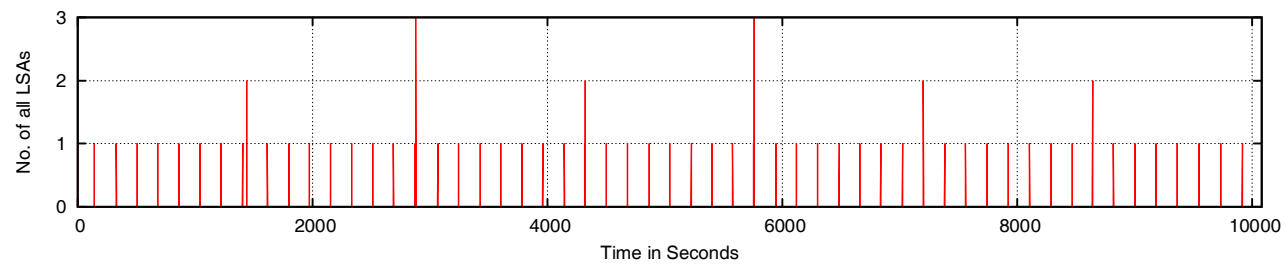

(a) Number of LSAs type-1 captured by rcs1

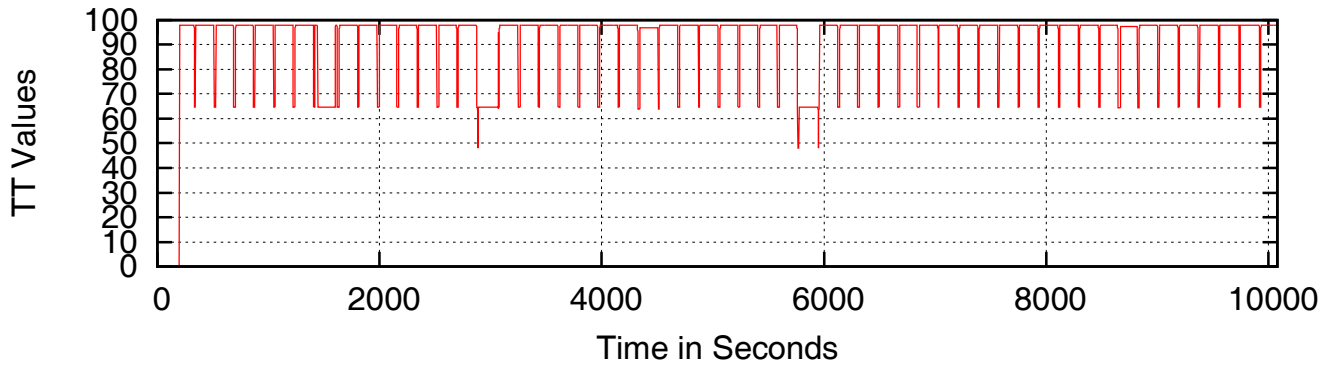

(b) TT measurement

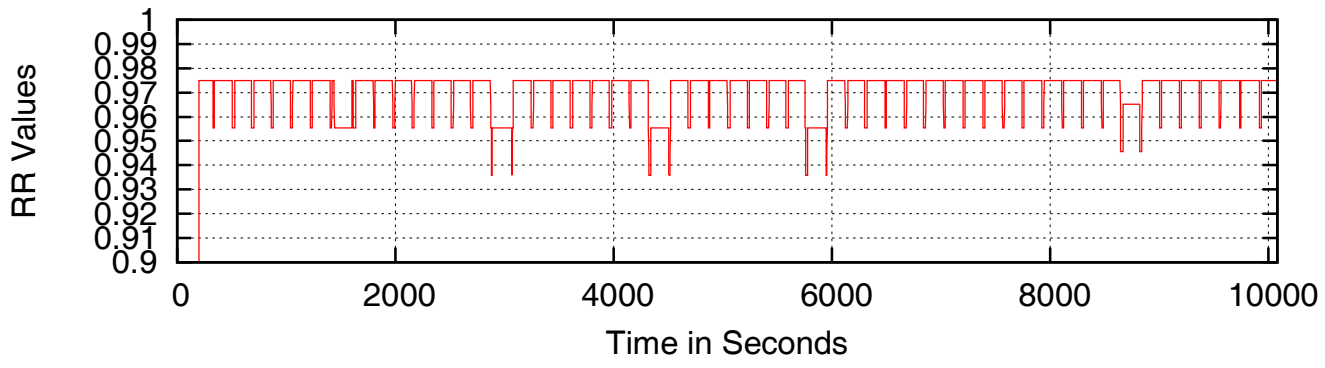

(c) RR measurement

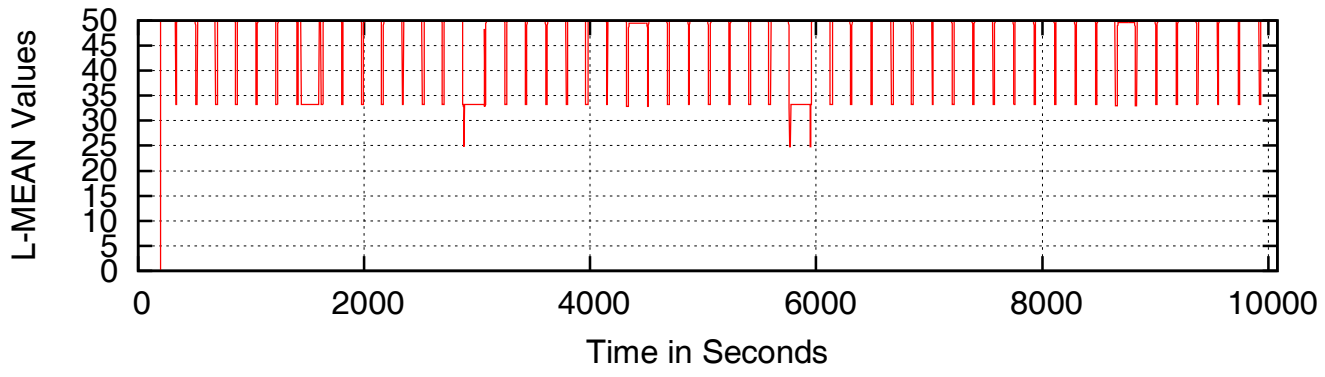

(d) L-MEAN measurement

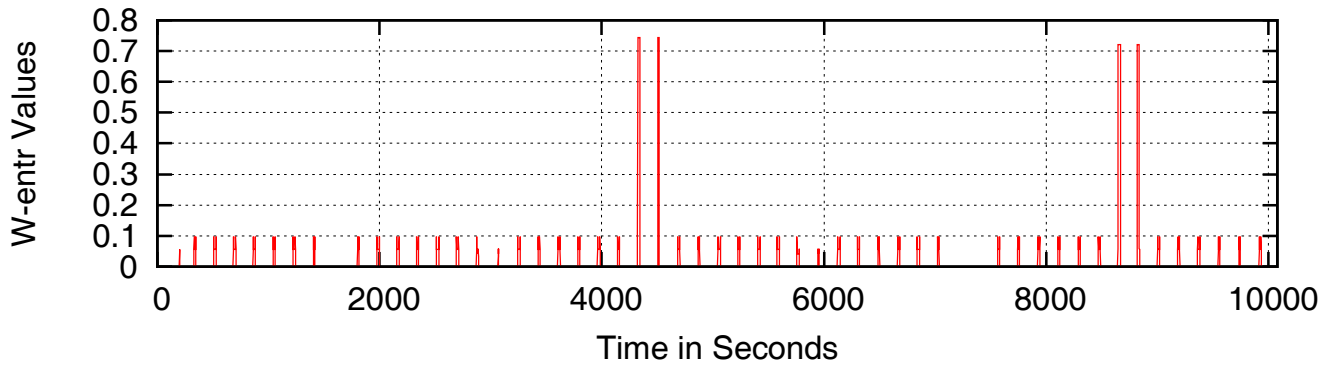

(e) W-entr measurement

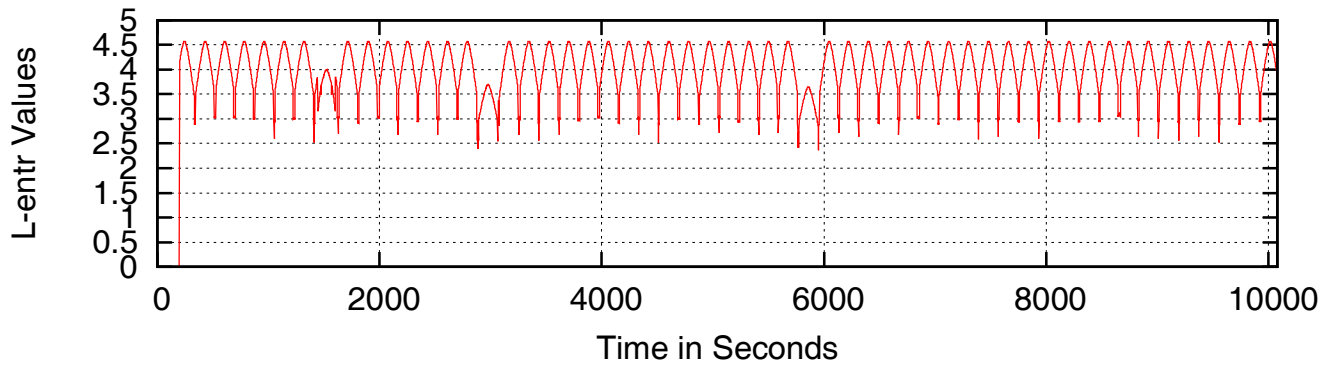

(f) L-entr measurement

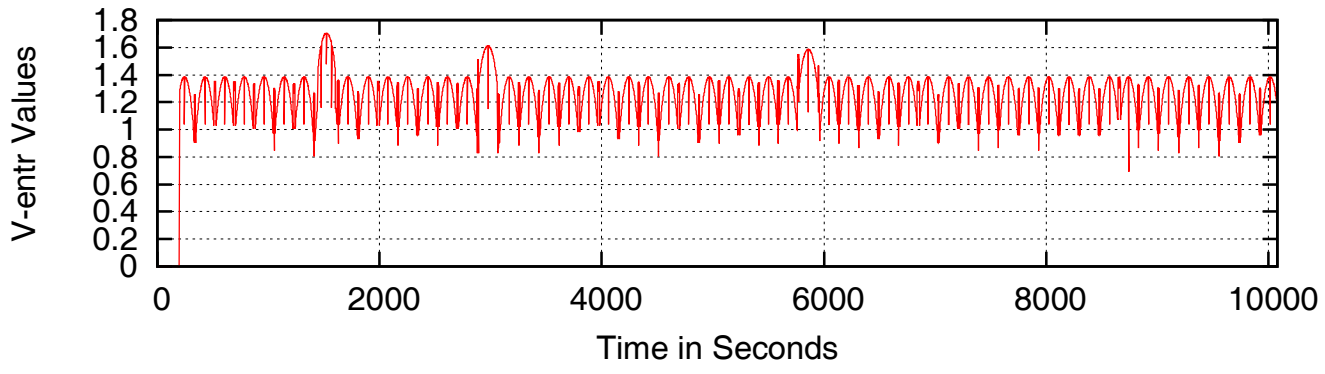

(g) V-entr measurement

Fig. 8. Number of LSAs type-1 advertised by abr1 and its corresponding values of RQA measurements

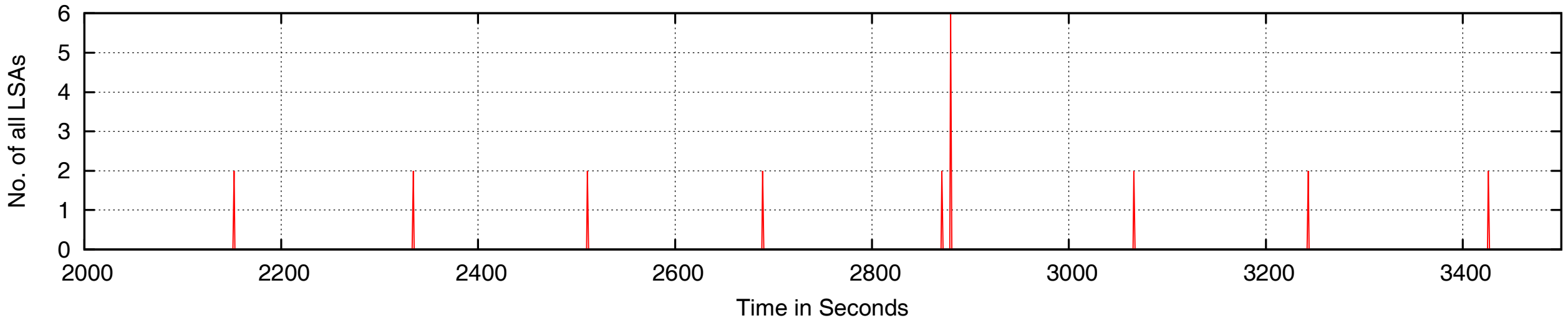

(a) Number of all LSAs during turning up eth0 of abr1

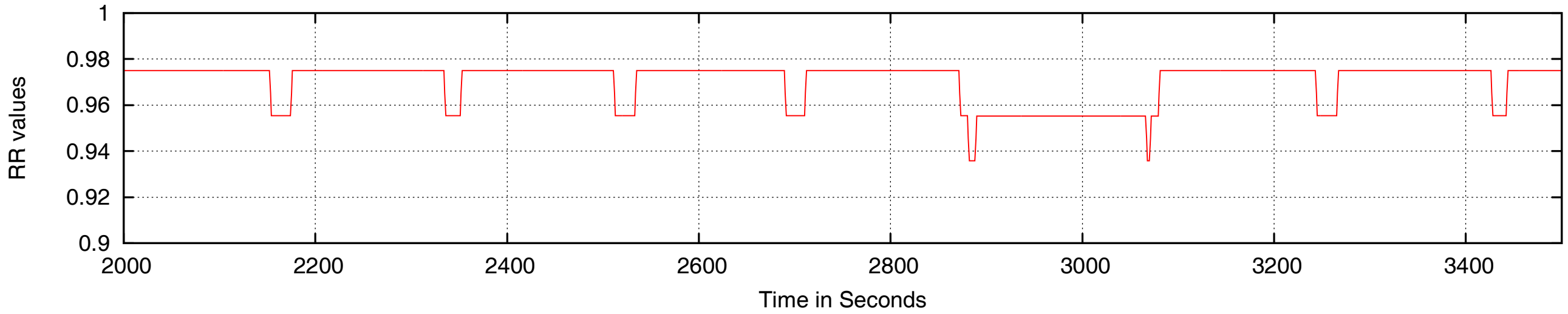

(b) Corresponding RR values for all LSAs during turning up eth0 of abr1

Fig. 9. An example of capability of RQA to rapidly identify OSPF failure

Once again, we use an automation script that introduces the three well known attacks. We introduced disguised, adjacency, and partitioning attacks at time stamps 2485, 5012, and 9532 seconds respectively. We assume r8 is the compromised router for disguised and partitioning attacks while host2 is the compromised host. Figure 10 shows the total number of all LSA updates advertised by r9 OSPF router and its corresponding measurements of RQA. In this example, we can see that W-entr measurement shows a significant change during the attackers time while in hardware failure RR was the significant measurement that identifies the failures. This observation can be used for further research in term of differentiate between type of anomalies.

## VI. Conclusions

OSPF is an intra-domain routing protocol that has been widely deployed for the last two decades. Different types of attacks exploit OSPF vulnerability. LSA falsification is one of the most serious attacks that threaten OSPF protocol. Rapid identification OSPF anomalies helps to mitigate the effect of OSPF from worse consequences such as route loop and black hole. In this paper, we show that RQA can be used to identify such anomalies. We evaluate the capability of RQA using data collected from a controlled testbed where different types of OSPF anomalies were introduced. Our evaluation shows that RQA is able to rapidly identify OSPF anomalies.

Our feature work will involve evaluate the capability of RQA to identify anomalies in a series of OSPF traffic originated by all OSPF routers. Our further research includes investigation the capability of RQA to differentiate between different OSPF anomalies.

## Acknowledgements

This work was conducted in partnership with the Communication Networks Research Science and Technology Capability of Defence Science and Technology Group within the Australian Department of Defence and the Cyber-Physical Systems Research Program of Data61/CSIRO, through the Next Generation Technologies Program.

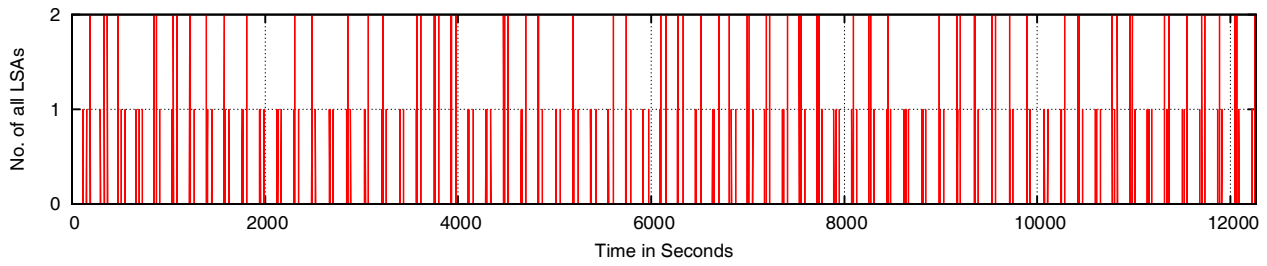

(a) Number of LSAs type-1 captured by rcs1

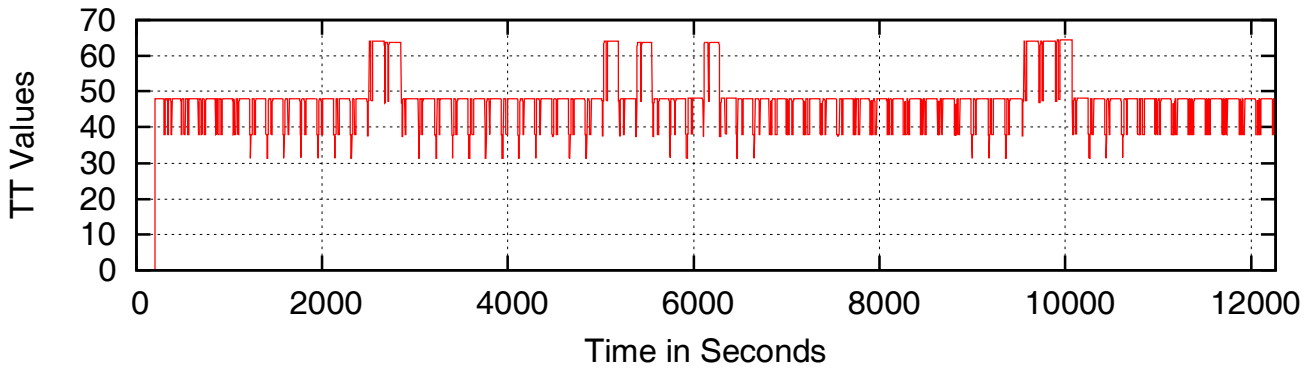

(b) TT measurement

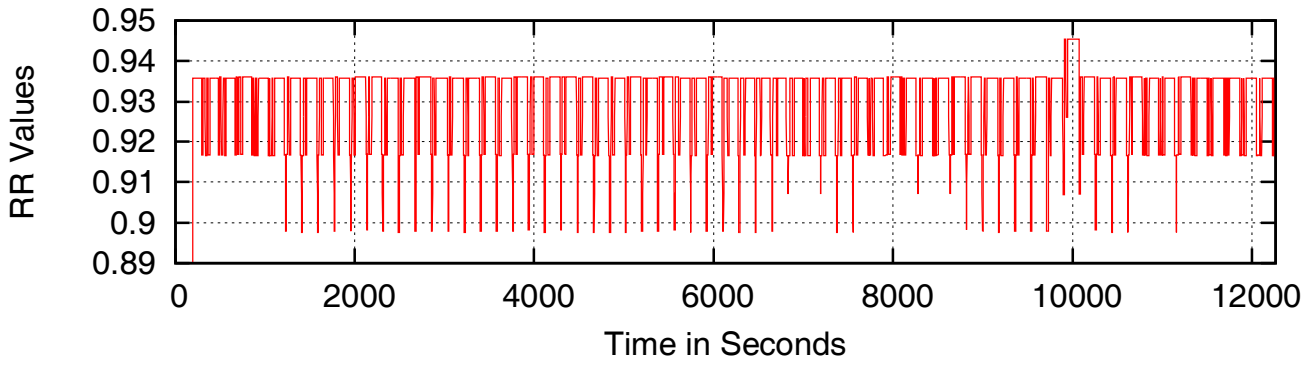

(c) RR measurement

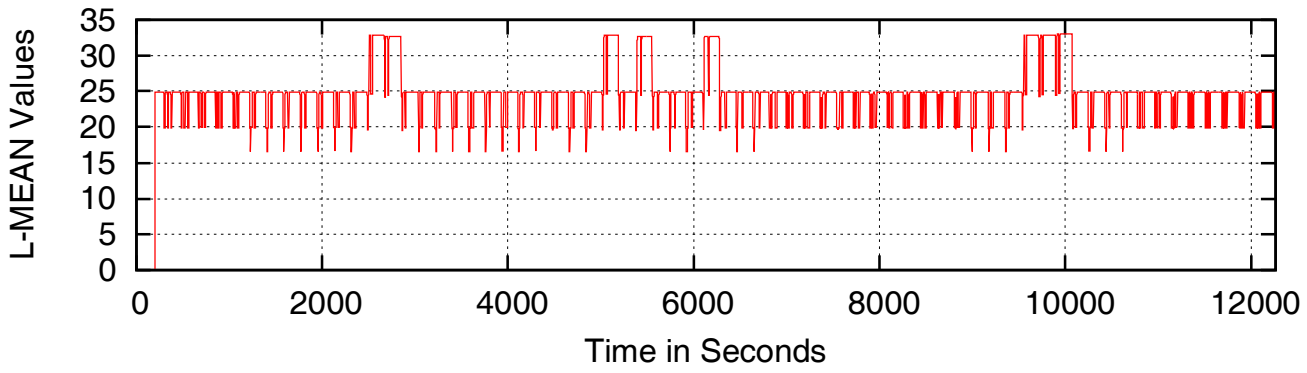

(d) L-MEAN measurement

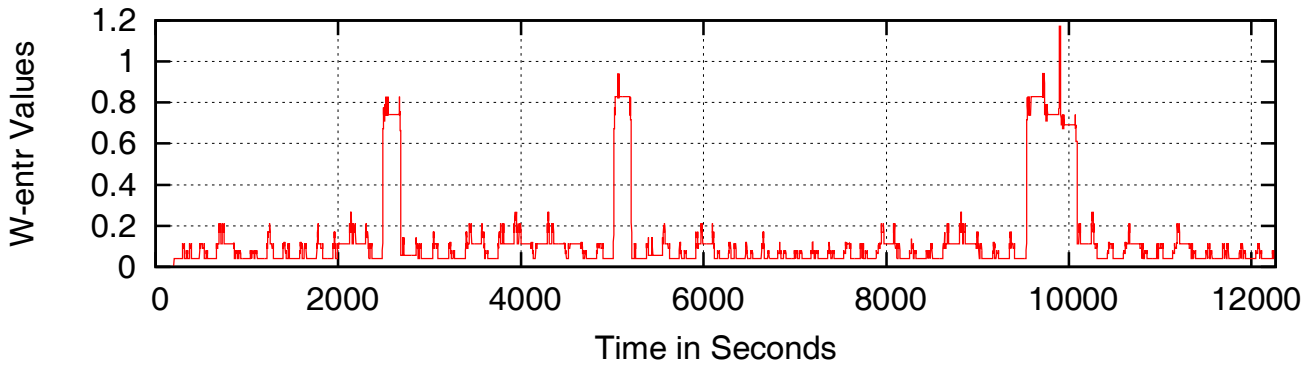

(e) W-entr measurement

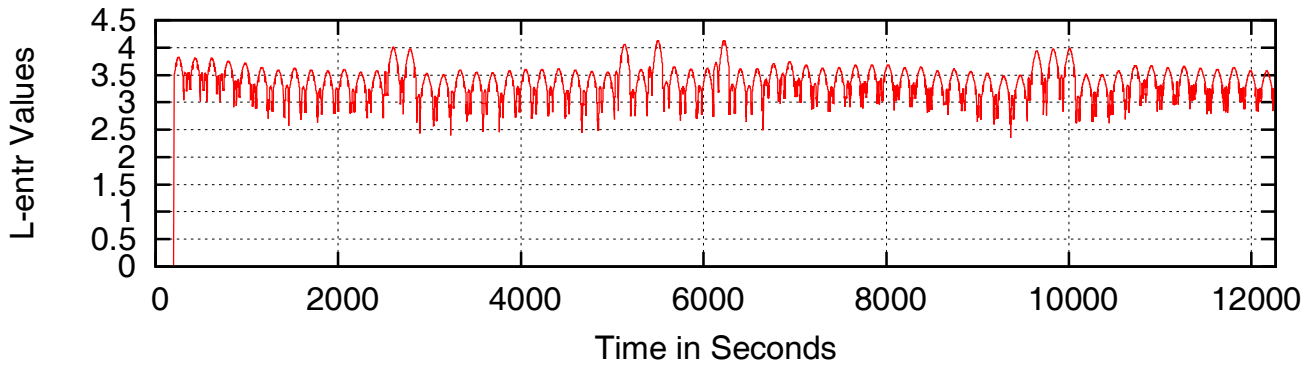

(f) L-entr measurement

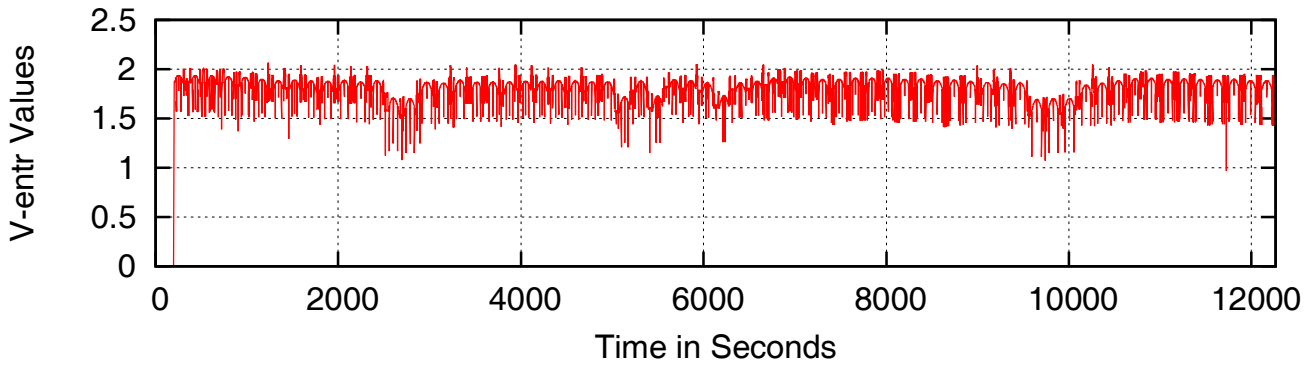

(g) V-entr measurement

Fig. 10. Number of all LSAs captured and its corresponding values of RQA measurements